\documentclass[runningheads]{llncs}
\usepackage[T1]{fontenc}
\usepackage{graphicx}

\usepackage [english]{babel}
\usepackage [autostyle, english = american]{csquotes}
\MakeOuterQuote{"}

\begin{document}
\title{Playing with Data: An Augmented Reality Approach to Interact with Visualizations of Industrial Process Tomography}
%
%
%
\author{Yuchong Zhang\inst{1}\orcidID{0000-0003-1804-6296} \and
Yueming Xuan\inst{1}\and \\
Rahul Yadav\inst{2}\orcidID{0000-0003-3822-3800} \and
Adel Omrani \inst{3} \orcidID{0000-0002-9602-9048} \and
Morten Fjeld \inst{1,4} \orcidID{0000-0002-9562-5147}}
\authorrunning{Yuchong Zhang et al.}
%
\institute{Chalmers University of Technology, Gothenburg, Sweden \\
\email{yuchong@chalmers.se} \and
University of Eastern Finland, Kuopio, Finland \and
Karlsruhe Institute of Technology, Karlsruhe, Germany \and
University of Bergen, Bergen, Norway\\
\email{fjeld@chalmers.se}, \email{Morten.Fjeld@uib.no}}
\maketitle              

\begin{figure}[tb]
 \centering
  \centering
  \includegraphics[width=\columnwidth]{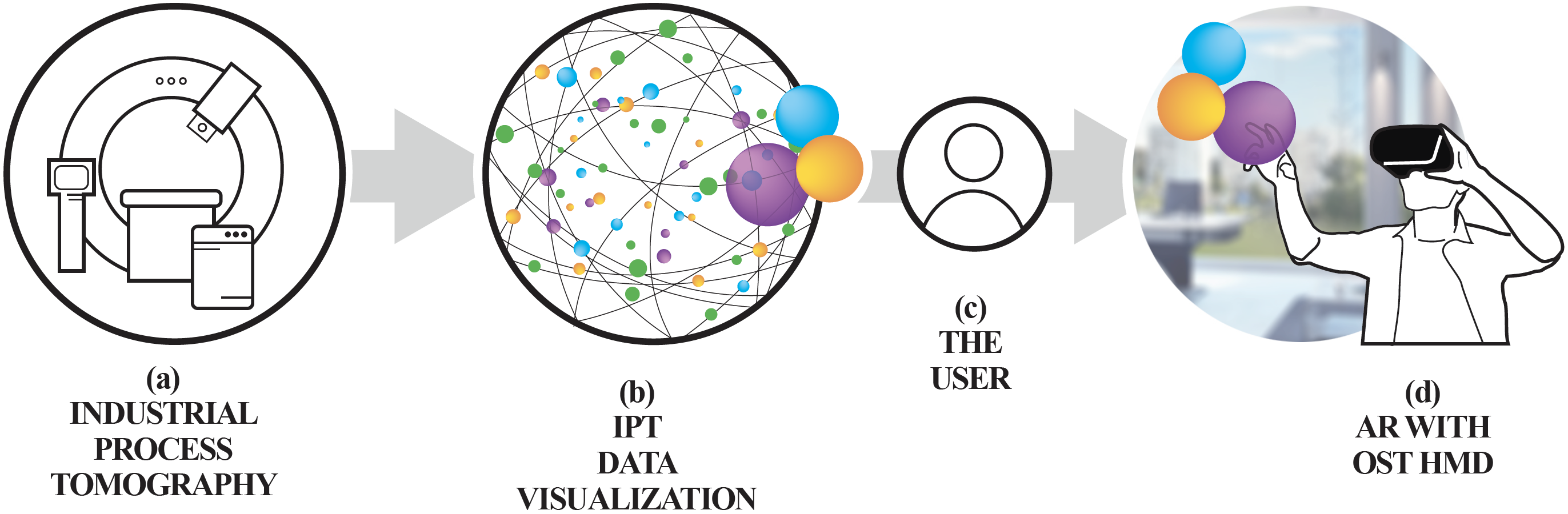}
  \caption{Conceptualization of our proposed AR approach for IPT visualization analysis. $a)$: The specialized IPT controlled industrial process is implemented in a confined environment. $b)$: The IPT data visualizations originated from different processes are imported and displayed in OST HMD AR. $c)$: Users engage with relevant data analysis. $d)$: User equipped with the proposed AR environment by an OST HMD interacting with the visualizations for further analysis.}
  \label{fig:teaser}
  \end{figure}

\begin{abstract}
Industrial process tomography (IPT) is a specialized imaging technique widely used in industrial scenarios for process supervision and control. Today, augmented/mixed reality (AR/MR) is increasingly being adopted in many industrial occasions, even though there is still an obvious gap when it comes to IPT. To bridge this gap, we propose the first systematic AR approach using optical see-through (OST) head mounted displays (HMDs) with comparative evaluation for domain users towards IPT visualization analysis. The proof-of-concept was demonstrated by a within-subject user study ($n$=20) with counterbalancing design. Both qualitative and quantitative measurements were investigated. The results showed that our AR approach outperformed conventional settings for IPT data visualization analysis in bringing higher understandability, reduced task completion time, lower error rates for domain tasks, increased usability with enhanced user experience, and a better recommendation level. We summarize the findings and suggest future research directions for benefiting IPT users with AR/MR.

\end{abstract}
\section{Introduction}
\label{intro}

Industrial process tomography (IPT) (Figure ~\ref{fig:teaser}.$a$) is a dedicated and non-invasive imaging technique which is pervasively used in manufacturing scenarios for process monitoring or integrated control \cite{ismail2005tomography,tapp2003chemical,nolet2012seismic,primrose2015application,plaskowski1995imaging,rao2022monitoring,zhang2020automated,zhang2020condition}. It is an effective mechanism to extract complex data, visualize it and interpret it for domain users \cite{hampel2022review,beck2012process,yao2017application,zhang2021affective,rao2022monitoring}. Due to the speciality and professionality of IPT, some rising technologies are therefore harnessed for both experts and laymen to perform complex tomographic data analysis to improve their productivity and efficiency. Through augmented reality/mixed reality (AR/MR) \cite{speicher2019mixed}, a now highly popular technique which superposes virtual objected into the real world \cite{leebmann2004augmented,liu112018application,tainaka2020guideline,zhang2022initial}, it is possible to provide the digital information extracted from intricate tomographic data to the front view of users. As a representative human computer interaction technology, AR consolidates the interplay between digital objects and human subjects with immersion based on a powerful and interactive AR visualization modality \cite{kalkofen2008comprehensible,tonnis2005experimental,pierdicca2015making}. Such usage has still not become mainstream in most industries but a moderate number of research projects have successfully demonstrated its value \cite{de2020survey}. Yet, wide adoption of AR within IPT is scarce due to the lack of technology fusion, and challenges such as hardware limitation and ease of use \cite{zhang2023need}.

AR features an exceptional characteristic that establishes immersive experiences where users intimately interact with the virtual object floating in front of their field of view \cite{azuma1997survey,billinghurst2015survey,zhang2022site}. With the rapid development of hardware technology, Wearable head-mounted display (HMD) AR has become widespread in many contexts since it enables interaction between physical objects coupled with digital information using human hands \cite{kahl2021investigation}. Under this circumstance, manipulation tasks that require interaction with the virtual objects can be achieved more accurately and quickly than when using conventional tangible interfaces \cite{besanccon2017mouse,kress2014segmentation}. Optical see-through (OST) HMDs occupy the majority of contemporary AR contexts due to their ability to use optical elements to superimpose the real world with virtual augmentation \cite{peillard2020can}. This differs from video see-though AR where the camera calibrating is realized by handling the captured image: in an OST AR system, the final camera is the human eye, and thus there is no captured camera imagery to process \cite{khan2021measuring}. With the ability of displaying manifold data in a more perceivable and interactive way \cite{heemsbergen2021conceptualising,dubois2001consistency}, applying AR to generate interactive visualizations for user-oriented tasks is often the preferred choice, and not only for industrial applications \cite{masood2020adopting,fite2011there,ong2013virtual,mourtzis2020augmented,alves2022comparing}. More specifically, it has been proven that OST AR can insert virtual information into the physical environment with visual enhancement for human observers to control, thus improving the performance of information-dense domain tasks and further adding to its appeal to industrial practitioners \cite{dunn2020stimulating,henderson2011augmented}.

While industrial AR has attracted considerable attention among researchers and engineers, the gap between this emerging technique and actual IPT scenarios still remains \cite{zhang2023need,zhang2021novel}. Recently, some practitioners have acknowledged the high value and potential of AR to be adopted in a variety of IPT-related situations. The combination of IPT and AR is seen as a thriving methodology since AR has the capacity to handle sophisticated 2D/3D data in an immersive, interactive manner, under the premise that IPT is used for monitoring and controlling confined processes. When combined, AR and IPT have a strong potential for visualization and interpretation of complex raw data \cite{nowak2019towards,chen2016using,nowak2021augmented}. The visualization supportablity and interactivity of AR is well justified in some IPT-controlled processes, including visualizing fluid mixture for stirred chemical reactors \cite{mann2001augmented} and providing in-situ collaborative analysis in remote locations regarding numerical data \cite{nowak2019towards}. Nonetheless, even though plenty of mobility has been brought about, there is no research into the essential interaction between IPT data visualizations and domain related users with the aid of OST AR.
 
In this paper, we offer a novel AR approach which concentrates on the interaction between data visualization from IPT (Figure ~\ref{fig:teaser}.$b$) and pertinent users (Figure ~\ref{fig:teaser}.$c$) to allow them to easily observe and manipulate the visualizations with high immersion. The target domain users are those who are in need of getting involved with IPT to any extent. We propose deploying AR/MR applications to tackle the practical problems stemming from the specific area (interactive IPT visualization analysis). The main advantage of the proposed methodology is that it initiates, to our knowledge, the first mechanism for furnishing IPT users with OST HMD AR with comparative evaluation to communicate with informative visualizations for better comprehension of industrial processes. The AR system employs Microsoft HoloLens 2; one of the representative OST HMDs (Figure ~\ref{fig:teaser}.$d$), as the fundamental supplying equipment to create the AR/MR environments. The source data derived from the IPT supported industrial processes is always formidable to understand, so needs straightforward and precise patterns to be visualized and interpreted. Our proposed approach provides a systematic framework which adopts accurate 2D/3D IPT data visualizations and supplies interactive analysis that is comprehensible to domain users regardless of their IPT experience. We carried out a comparative study to demonstrate the superiority of our AR approach on bringing interactive visualization surroundings as well as eliciting better contextual awareness. We envision our AR approach to benefit any areas where users deal with IPT visualization analysis. The main contributions of this paper are as follows:

\begin{itemize}
    \item Proposing a novel AR approach for domain users to perform contextual IPT data visualization analysis using OST HMDs, with better understandability, task performance and user experience.  
    \item Designing and implementing a comparative study with user testing to prove the effectiveness of our proposed AR approach compared to the conventional method and acquiring early-stage feedback.
\end{itemize}

The structure of this paper is as follows: The background and motivation are narrated Section ~\ref{intro}. Section ~\ref{rw} presents state-of-the-art related works of using AR with IPT for interactive visualization in related contexts. Section ~\ref{meth} comprehensively presents of the proposed AR approach, including system overview, contextual stimulus, and the  apparatuses. Section ~\ref{ed} offers experimental design and implementation for evaluation. Section ~\ref{dis} discusses insights and limitations. Finally, Section ~\ref{cnf} draws concluding remarks and future work.

\section{Related Work}
\label{rw}
AR/MR and derivative techniques are gradually appearing in more industrial contexts \cite{bottani2019augmented,regenbrecht2005augmented} due to AR offering computer-generated virtual and context/task related information in-situ for users \cite{jasche2021comparison}. The interactivity derived from AR technology regarding virtual visualizations in various industrial application scenes has particularly been advocated by researchers over the past few years.  

\subsection{Augmented Reality Visualizations for Industry}
Over a decade ago, Bruno et al. \cite{bruno2006visualization} developed an AR application named VTK4AR, featuring that functionality which uses AR to scientifically visualize industrial engineering data for potential manipulation of the dataset. The many requirements of applying AR within industrial applications have been summarized by Lorenz et al. \cite{lorenz2018industrial} as they enumerate the user, technical, and environmental requirements. Mourtzis et al. \cite{mourtzis2020augmented} proposed a methodology to visualize and display the industrial production scheduling and monitoring data by using AR, empowering people to supervise and interact with the incorporated components. In industrial manufacturing, a comparative study conducted by Alves et al. \cite{alves2022comparing}, demonstrated that users who replied on AR based visualizations obtained better results in physical and mental demand in production assembly procedures. Even more specifically, B\"{u}ttner et al. \cite{buttner2016using} identified that using in-situ projection based spatial AR resulted in a significantly shorter task completion time in industrial manual assembly. They designed two assisting systems and proved that the projection based AR outperformed HMD based AR in ease of use and helpfulness in their context. Avalle et al. \cite{avalle2019augmented} proposed an adaptive AR system to visualize the industrial robotic faults through the HMD devices. They developed an adaptive modality where virtual metaphors were used for evoking robot faults, obtaining high effectiveness in empowering users to recognize contextual faults in less time. Satkowski et al. \cite{satkowski2021investigating} investigated the influence of the physical environments on the 2D visualizations in AR, suggesting that the real world has no noticeable impact on the perception of AR visualizations.

\subsection{Augmented Reality for Industrial Process Tomography}
Even though the volume of intersection research between AR and IPT is not substantial, other researchers have been investigating diverse pipelines of applying AR to generate interactive and effective visualizations for complex tomographic process data. Dating back to 2001, Mann et al. \cite{mann2001augmented} exploited an AR application to visualize the solid-fluid mixture in a 5D way in stirred chemical reactors operated by one breed of IPT--electrical resistivity tomography (ERT). A few years later, Stanley et al. \cite{stanley2005interrogation} directed the first study of applying ERT to a precipitation reaction process by using AR visualization to display images shown in a suitable format. Zhang et al. \cite{zhang2023need} conducted a need-finding study investigating the current status and prospective challenges of AR in IPT. They pointed out that there is a great potential of deploying cutting-edge AR technique among IPT practitioners. A new solution to visualize IPT data, leading to collaborative analysis in remote locations, was proposed by Nowak et al. \cite{nowak2019towards}. More specifically, their team also explored a more in-depth AR system with a more advanced but still preliminary prototype which created an entire 3D environment for users to interact with the information visualizations characterizing the workflow of IPT with better immersion \cite{nowak2021augmented}. The OST HMD AR was satisfactorily adopted in the experiment they conducted but there were no specific interaction and evaluation included. Later, Zhang et al. \cite{zhang2021supporting} formulated a novel system to generalize IPT within the context of mobile AR, and directed a proof-of-concept implementation where an AR framework was developed to support volumetric visualization of IPT, especially to yield high mobility for users but without involving any practical user testing. Sobiech et al. \cite{sobiech2022exploratory} conducted an exploratory study on user gestural interactions with general IPT 3D data by involving HMD AR, affirming the initial adoption of AR in interacting with IPT data, however, no comparison to conventional tools and no structured user study were engaged either. At present, there is to our knowledge no systematic research into framing the use of OST HMD AR for IPT data visualization analysis and comparing it with the conventional tools through constructive user studies.

\begin{figure}[tb]
 \centering
 \includegraphics[width=.75\columnwidth]{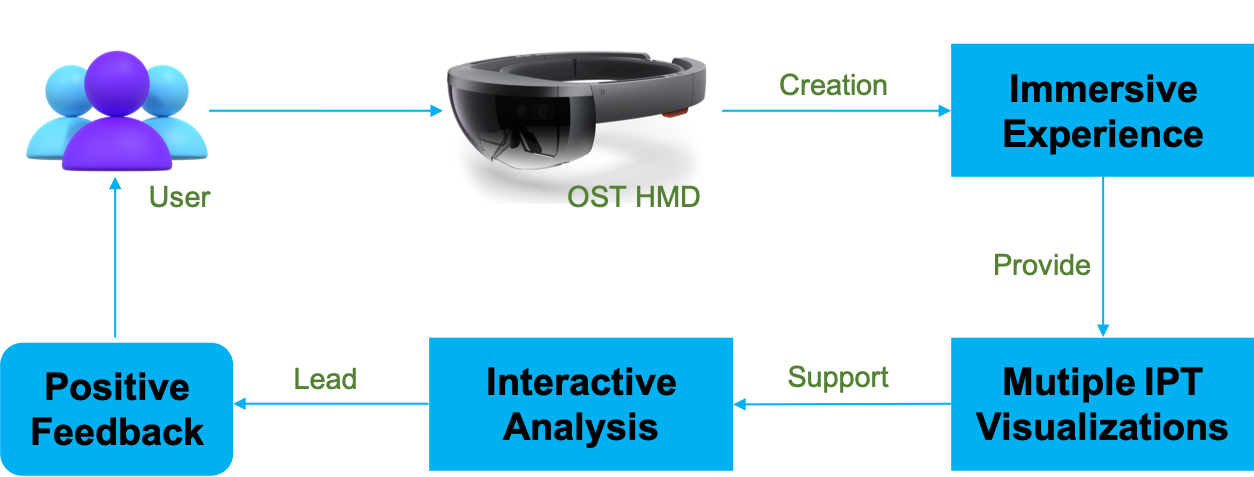}
 \caption{The block diagram of our proposed AR approach. The user starts the procedure by wearing the OST HMD. The immersive experience is then created, which accommodates the multiple IPT data visualizations for supporting interactive analysis. The user feedback is then obtained.}
 \label{fig:sys}
\end{figure}

\section{Methodology}
\label{meth}

\subsection{System Overview}
Our AR system is designed as an interactive approach to accommodate users with immersion to interact with different visualizations originated from IPT monitored contexts. A schematic diagram of the system is shown in Figure ~\ref{fig:sys}. The practical embodiment of our approach is an AR application with switchable options for different modalities of visualizations being placed accurately in front of human eyes for desired immersion \cite{ens2021grand}. Users launch the AR app after putting on the OST HMD. This enables them to manipulate visualizations, generated from IPT supervised processes, with their hands. As displayed in Figure ~\ref{fig:user_play}, the IPT visualization is cast as a floating object in our AR context, while users are supported to intimately drag (grabbing the visualization to force the motion as it deviates from the original position), rotate (grabbing the edges of the visualization to turn or spin it around its central point), zoom in/out (grabbing the edges of the visualization to adjust it's level of magnification), and execute other physical manipulations (moving, flipping, aligning, etc) in order to get deeper understanding of the data and perform further analysis. The virtual objects in front of the users' eyes are at a proper distance and completely visible within the field of view of the OST HMD, so they can view all the content shown. The provided AR interaction formulated for users towards these intricate tomographic data visualizations is deemed to be the most striking attribute, and has not been indicated in any previous IPT scenarios. Furthermore, a user feedback component is involved in this approach to appropriately appraise this AR system. We designed and implemented a comparative user study to evaluate our proposed methodology and gain early-stage feedback, which is presented in Section ~\ref{ed}.    

\subsection{Stimulus}
\begin{figure}[tb]
 \centering
 \includegraphics[width=.6\columnwidth]{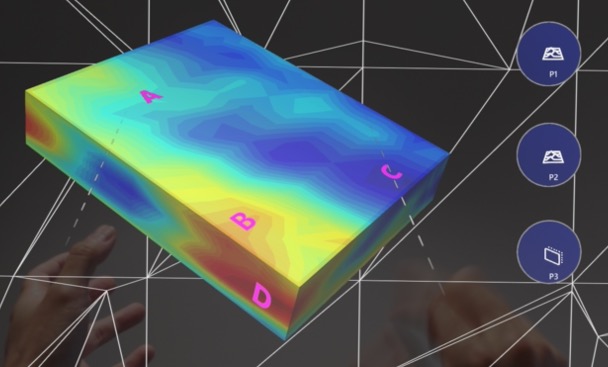}
 \caption[]{An example scene of the user interacting with one of the three visualizations derived from IPT in AR environment. Different colors and the annotations represent different moisture levels of the polymer material after the microwave drying process. Virtual buttons \protect\footnotemark for switching to alternative visualizations are located on the right side.}
 \label{fig:user_play}
\end{figure}

The contextual data used in this study was acquired from a microwave drying process for polymer materials monitored by a specific genre of IPT -- microwave tomography (MWT). This imaging modality was applied to detect the moisture levels and distribution of the polymer material through specific tomography imaging algorithms \cite{omrani2022multistatic}. Three different visualizations were employed in this study, obtained from three dissimilar MWT drying processes including two in 3D and one in 2D, as 3D figures incorporate more information regarding the polymer materials used in the process. The three visualizations were positioned individually in three interchangeable interfaces with three virtual buttons arranged on the right-hand side to switch to another. An example including one 3D visualization and the buttons are shown in Figure ~\ref{fig:user_play}. As displayed, different moisture levels are rendered with distinct colors and marked A, B, C and D, denoting low moisture area, moderate moisture, the dry part, and the high moisture area respectively. The annotations were created based on the physical understanding by the domain expert tightly correlated to the process. Nevertheless, it is critical for users who intend to deal with IPT related analysis, regardless of their expertise, to fully understand the contextual information for in-depth visualization analysis.

\footnotetext{The darker background and white grid lines were only visible during the app activation and the screenshot capture process, but not visible during the experiment.}

\subsection{Apparatus}
For the environmental configuration of our AR system, we used Microsoft HoloLens 2 glasses; the specialized OST HMD with one dedicated application developed on it. This HMD weighs 566 grams with a built-in battery, and has see-through holographic lenses, a SoC Qualcomm Snapdragon 850, and a second-generation custom-built holographic processing unit. All of these characteristics enable the user to wear it freely and comfortably. The AR application was developed with Unity3D 2020.3.20f1 and Mixed Reality Toolkit 2.7 (MRTK). MRTK is an open-source development kit for creating MR applications for Microsoft HoloLens and Windows MR devices. It provides a number of fundamental components to build user interfaces and interactions in mixed reality settings. The MRTK components "NearInteractionTouchable" and "NearInteractionGrabbable" were combined in this application to provide hand tracking and touch input on the visualization figures. Additionally, the offered "ObjectManipulator" component was utilized to allow users to grab and move the figures with their hands, with gestures like translation, rotation, and scale enabled. Furthermore, three virtual buttons were created for switching between visualization figures using the "PressableRoundButton" component. For the conventional visualization analysis settings, a 16-inch MacBook Pro (3072*1920) with the operating system macOS Monterey version 12.3.1 was used to conduct the comparison study. The actual tool used for users to observe and manipulate the data visualizations in the conventional setting was Matlab\_R2021\_a (the common tool used in IPT related visualization and analysis \cite{omrani2022multistatic}) with visualization window.

\section{Experiment Design}
\label{ed}
A user study was conducted to find out whether our proposed AR approach, offering interactive and immersive experience for communicating with IPT visualizations, was superior to customary in-situ data analysis pertaining to supporting users' understandability and user experience. The main realization of this study was two different scenarios where the engaged participants encountered two different environments; our proposed AR approach to interact with the data representations with the aid of OST HMD; and the conventional setting using an ordinary computer with 2D screen and Matlab (only the visualization window was utilized for the experiment). Three tasks were designed towards IPT contextual visualization analysis for the participants to implement. We selected the within-subjects principle and examined all the participants in both scenarios, and their performance when dealing with the data visualizations by means of several specific domain tasks. To precisely investigate the feedback as well as mitigate the order effect from the displacement of the two environmental settings, we followed the counterbalancing design principle \cite{bradley1958complete}. We divided the participants into two groups (AR-first group equipped with the AR approach and baseline-first group equipped with conventional computer involving a 2D screen at the beginning) with ten people in each. The AR-first group was first equipped with our AR system to complete the tasks, then they were shifted to the conventional computer with 2D screen for the same tasks while the baseline-first group was treated in the reverse order. After task completion, each participant was directed to fill in a post-study questionnaire followed by a short interview session. The total study duration was 12--18 minutes. As aforementioned, the ultimate goal of the designed system was to assist domain users for better in-depth comprehension of the IPT data, allowing better visualization analysis through the AR technique. Therefore, the following hypotheses were made in our study regarding IPT contextual data visualization and task performing:

\begin{itemize}
    \item \textbf{H1:} Our proposed AR approach can obtain better understandability for users compared to using the conventional setting.
    \item \textbf{H2:} Our proposed AR approach can contribute to a lower task completion time and fewer task errors.
    \item \textbf{H3:} Our proposed AR approach has better usability for users to interact with IPT data visualizations.
    \item \textbf{H4:} Our proposed AR approach has a greater recommendation level than the conventional setting.
\end{itemize}

\subsection{Participants}
Twenty participants ($n$=20, 12 self-identified male and 8 self-identified female) aged between 24 and 41 ($M$ = 30.1, $SD$ = 4.78) were recruited by either e-mail or personal invitations at a local university. All the participants were proficient in dealing with most of the computer operations within the MacOS system. They were first asked about their familiarity and prior experience with AR in general as well as any previous exposure to IPT. Only two reported occasional experience with AR while four had previous experience with IPT. The proficiency of Matlab was not considered since the participants were merely required to conduct the tasks within the Matlab visualization window through simple operations without any coding procedures or data creation. Participants were then evenly and randomly divided into the AR-first group ($n$=10) and baseline-first group ($n$=10). None reported any kind of discomfort when wearing the HMD while each completely finished the whole study procedure, and none reported circumstances of mis-seeing any virtual objects through HoloLens 2. The study was conducted in a bright and spacious function room without any other distractors. We acted in strict accordance with the Covid-19 rules in each step of the study. Each participant was rewarded with a small gift for helping with the study. All the data collection conformed with the ethical guidelines of the local university where the study was carried out.

\subsection{Procedure}
All participants signed an informed consent at the beginning of each session, which stated that there would be no personal information collected and that they could quit the study at any time. They were also told that the sessions would be audio-recorded, but all recordings would be treated confidentially, and references would only be made in a purely anonymized form for scientific analysis. All participants had sufficient time to read the consent form and ask questions before signing it. Due to the high professionality and complexity of IPT, the data generated is usually difficult to interpret by outsiders, even some experts. Hence, we started our study with a concise but detailed introduction about the study background, including the fundamental schematic of IPT, the source of the data visualizations, and the basic information about the visualizations used in this study. Before we began with the actual study sessions, we calibrated the Microsoft HoloLens glasses for each participant by helping them follow the instructions from the built-in calibration functional module. A short pre-training session was implemented to get the participants familiarized and adapted to the HMD and the AR application used.

\begin{figure}[tb]
 \centering
 \includegraphics[width=\columnwidth]{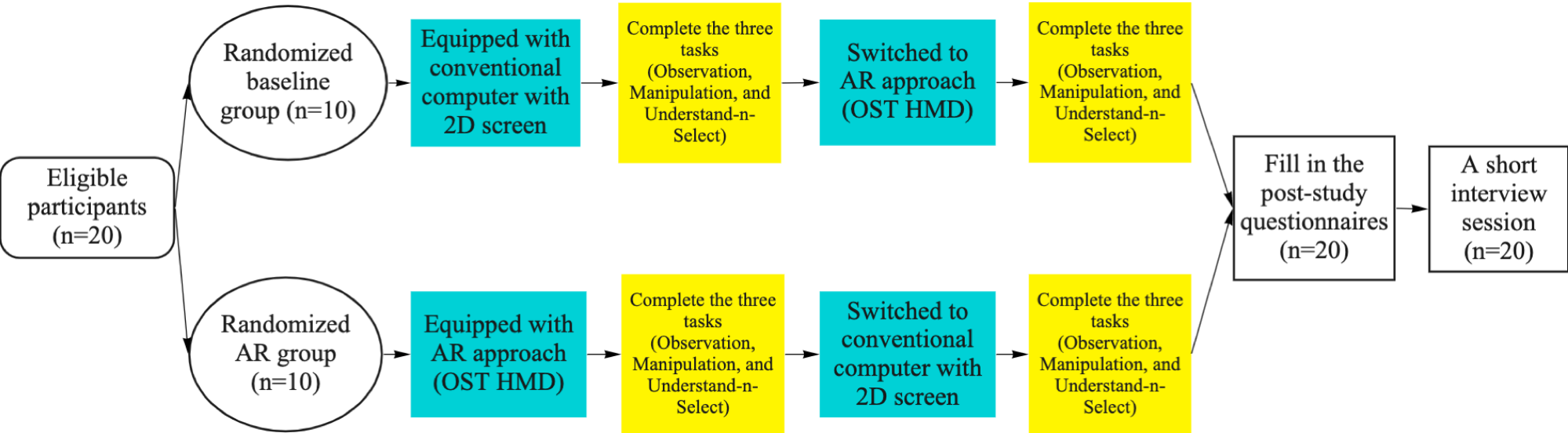}
 \caption{Flowchart of our user study with counterbalancing design.}
 \label{fig:us_flow}
\end{figure}

\begin{figure}[tb]
 \centering
 \includegraphics[width=\columnwidth]{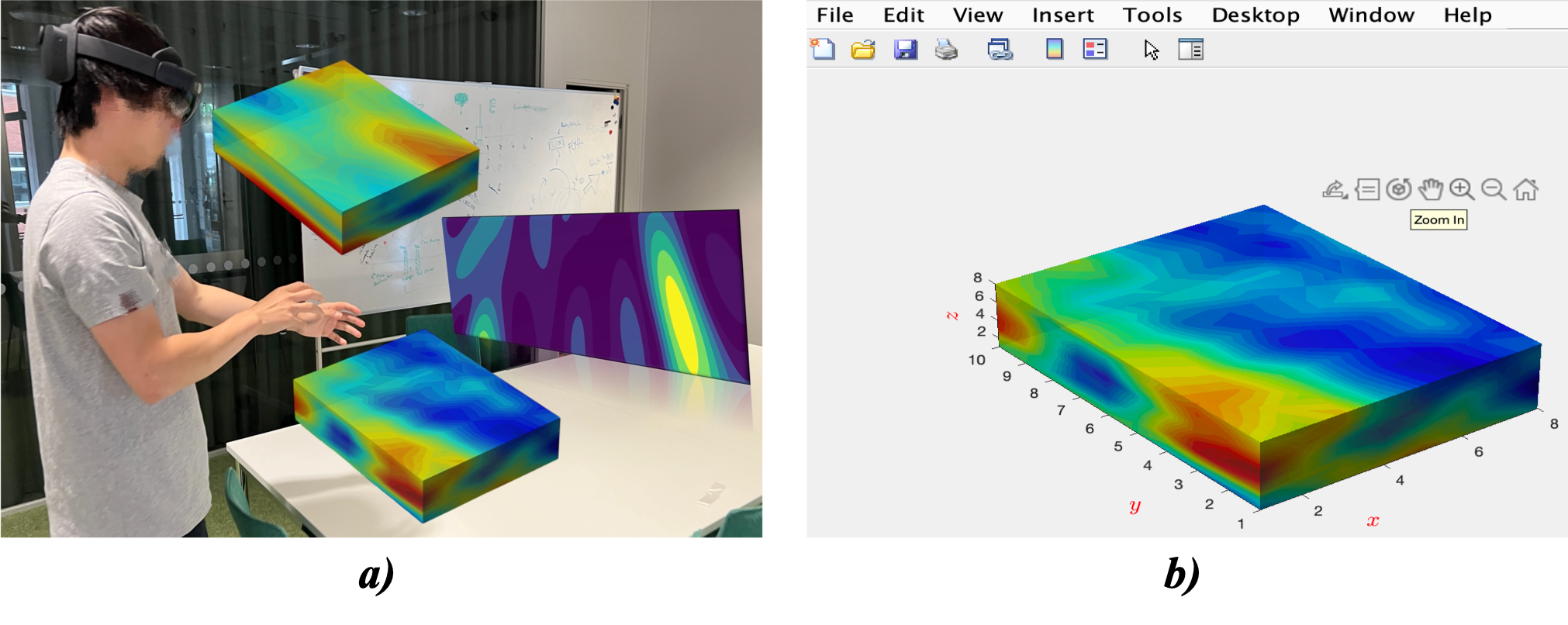}
 \caption{User study example scene of the two environmental settings. $a$: AR approach--the participant equipped with the OST HMD is interacting with the three IPT visualizations studied in this paper for data analysis. \textbf{Note:} the three visualizations are not placed together in one scene as the figure shows; they are placed individually in three different switchable interfaces in the AR app. $b$: Conventional computer with 2D screen--the IPT visualization is placed in a Matlab visualization window for participants to interact with. The 2D figure represents the surface information while 3D figures represent the volumetric information of the polymer materials used in the study.}
 \label{fig:user_play_3}
\end{figure}

The entire study procedure is displayed in Figure ~\ref{fig:us_flow}. Participants were required to complete three micro tasks regarding three selected figures to become familiarized with the IPT data visualizations as well as interacting with them in-situ. Each person had the identical opportunity to use both our AR system and the conventional tools to complete the tasks. Figure ~\ref{fig:user_play_3} shows an example scene of a participant being equipped with the two different environmental settings. They were also told they could convey real-time feedback in any form during the task implementation. The first task was called Observation, in which the participants were required to inspect the three visualizations (one 2D and two 3D figures), reading the introduction for five minutes but free to move to the next task of they felt ready. This was to familiarize them with IPT and also consolidate the information. After that they went onto the second task -- Manipulation. This required them to interact with the visualizations by dragging, rotating, and zooming in/out with either the AR context in the OST HMD or Matlab visualization window in the conventional computer. When manipulating with the AR glasses, they were free to walk around the visualizations to carry out the designated interactions. The last task was Understand-n-Select, where they were required to select four from a number of annotated pre-marked areas representing different moisture levels of the polymer materials on the three figures. The "Understand" component referred to that the participants had to understand the questions given related to IPT and annotations marked on the visualizations. The "Select" component implied that participants commenced doing practical tasks after understanding. All figures (2D and 3D) were shown identically in each environmental setting but the participants had to select four different areas by switching from one environment setting to another (they informed the authors the answers after the selection). The questions presented to the two settings are listed below.\\

\textbf{\textit{AR approach:}}
\begin{itemize}
    \item For Visualization 1, select the one area of moderate moisture level.
    \item For Visualization 2, the two areas of high and low moisture levels. 
    \item For Visualization 3, select the one area of low moisture level.
\end{itemize}
\textbf{\textit{Conventional setting:}} 
\begin{itemize}
    \item For Visualization 1, select the two areas of high moisture level and dry area.
    \item For Visualization 2, select the one area of moderate moisture level. 
    \item For Visualization 3, select the one area of dry area.
\end{itemize}
Participants were not allowed to refer to previous information when doing the tasks. We measured the task completion time (TCT) and the error rate of Understand-n-Select merely, since we believed it sufficient that it represented the hands-on task performance of the IPT visualization analysis. After task completion in both of the settings, a post-study questionnaire containing a few questions with quantitative scales to rate was completed. Finally, everyone was asked to attend a short interview session.

\subsection{Qualitative Results}
We compiled the results collected from the post interview sessions and real-time feedback of the participants during the study. The qualitative measurements generalized by a thematic analysis are presented in this section. The real-time feedback was conveyed spontaneously by the participants and recorded by the authors. For the post study interview, we asked the participants several subjective questions regarding what they liked and disliked about the tasks regarding our AR approach as well as the comparison between the two tested environmental settings. Concerning privacy and anonymity, we denoted the participants as P1 to P20 when recording the real-time feedback and encoding the interview answers. Based on the codes, we derived three themes: in-task understandability, interaction-based usability, and user experience.

\subsubsection{Theme one: In-Task Understandability}
It became apparent that the AR approach contributed to the understandability of the tasks' complex IPT data. Nearly half of the participants reported that the AR approach assisted in their understanding of the tasks, as it enabled convenient observation and offered useful information. For instance, when doing the tasks in the AR setting, P7 commented: \textit{"Now I understand the introduction better when playing with these [the figures] in AR"} and P15 remarked: \textit{"It's really helpful for me to understand the IPT by observing the visualizations in a 3D space"}. For similar reasons, P8 reported: \textit{"It was convenient to manipulate the visualizations near me, which was good for tasks"}. Furthermore, some owed their preference for the AR approach over conventional computers by articulating \textit{"3D objects present more information in the glasses"} (P5) and \textit{"I think I can see more content by rotating the figures with my hands by AR"} (P18). In contrast, none of the participants mentioned the understandability when using the conventional computer. This indicates that, as compared to traditional computers, the AR approach excels at making IPT data tasks more understandable, which supports our \textbf{H1}. Furthermore, a certain degree of evidence for our \textbf{H2} is shown by the fact that when the participants had a better knowledge of the IPT data, they were less likely to commit errors in task completion.

\subsubsection{Theme two: Interaction-Based Usability}
Usability of high interactivity is the most frequently mentioned aspect of the AR approach. Almost all the participants commented that the approach presented high usability in an immersive, interactive manner, whereas they deemed the interactivity of the conventional interface undesirable: e.g., \textit{"it was difficult to rotate and amplify them [the figures] in Matlab window especially when I wanted deeper observation, but it was very convenient to observe them [the figures] in AR"} (P3), and \textit{"It was nearly impossible to manipulate the figures in the computer. It felt nice to interact with an object in 3D"} (P5). Such remarks pointed out the manipulation of quality enabled by the AR approach. In essence, they helped confirm \textbf{H3} by acknowledging the approach's better usability for user interaction with the data visualizations than conventional computers. Moreover, the participants provided further explanations for the high usability perceived, as they gave credit to intuitive hand gestures and immersive 3D space distance. As for hand gestures, for example, P4 reported: \textit{"To use my hands to manipulate the figures in 3D is preferable"}. P9 reported: \textit{"It felt very intuitive to move the objects [the figures], and zoom in becomes easier"}. In addition, some who had visualization-related experience in Matlab discovered: \textit{"It was easy to switch to another figure in AR by my hands. There is no need to generate different codes as I usually did in Matlab"} (P10). When it comes to space distance, some found that the close distance from the 3D objects to them contributed to the interaction and usability, and remarked that \textit{"In AR, it was amazing to see the figures can be manipulated in desired way. I could make them [the figures] closer to me and see them more clearly"} (P4), and that \textit{"The interaction is really nice. It was cool to closely interact with the figures"} (P8). Comparatively, P14 affirmed: \textit{"It's super hard to zoom in the figures and it's impossible to move them in Matlab visualization window since the positions are fixed"}. To summarize, the AR approach achieved its satisfying usability mainly by allowing for immersive interaction, which enabled participants to manipulate the data visualizations in the 3D space using their hands.

\subsubsection{Theme three: User Experience}
Finally comes the theme of user experience offered by the AR approach, which most participants found to be favorable. Some described their use experience as pleasant: e.g., \textit{"The interaction with the figures in AR was adorable. It made me feel pleasant"} (P2), \textit{"It felt nice to interact with them [the figures] in 3D"} (P8), and \textit{"It was amazing to see AR with interaction"} (P9). Some discovered that \textit{"The immersion was really nice"} (P7), and that \textit{"The 3D visualizations made me feel real"} (P3). In addition, some deemed the task completion process enjoyable, because they \textit{"had fun exploring 3D space"} (P1), \textit{"The interaction in AR was interesting"} (P5), and \textit{"The interactive method in AR was easy to use. I enjoyed exploring the objects [the figures] there."} (P9). It can be seen that most participants gained good user experience when interacting with and explore the data visualizations immersively. Conversely, most participants expressed negative attitudes towards performing the tasks in the conventional computer. For example, P1 reported that \textit{"I could not do it properly in Matlab, but AR gave me gunny experience"}, P4 reported that \textit{"It was nearly impossible to manipulate the figures in the computer"}, and P8 reported that \textit{"It was horrible to interact with the figures in the computer"}. In other words, the AR approach has the potential to provide more enjoyable experiences than those frustrating ones experienced with conventional computers, which addresses \textbf{H3} to some extent.



\subsection{Quantitative Results}
We present our quantitative measurements here. The independent variables were identified as the two different environmental settings. From the study process and post-study questionnaire, five metrics -- TCT, understandablity, error rate, usability, and recommendation level employed as dependent variables were investigated. The TCT was merely measured from Understand-n-Select since this task is representative for practical IPT visualization analysis. To test the understandability, a 7-point Likert scale was adopted for participants to rate the level of the two environmental settings helping with understanding complex IPT data. The error was also exclusively collected from the third task when users were supposed to select the designated areas. The largest number of errors was four in each environmental setting. Every error, if one occurred, was noted during the study and the error rate was thereby calculated by the authors. The system usability scale (SUS) \cite{bangor2008empirical} was harnessed to quantify the usability. We uniquely investigated the recommendation level by asking the participants the extent (the 7-point Likert scale) of recommending the two environmental settings. We statistically analyzed the collected results to identify any significance among the four metrics evaluated. Normality was checked as a prerequisite.

\begin{figure*}[tb]
 \centering
 \includegraphics[width=\textwidth]{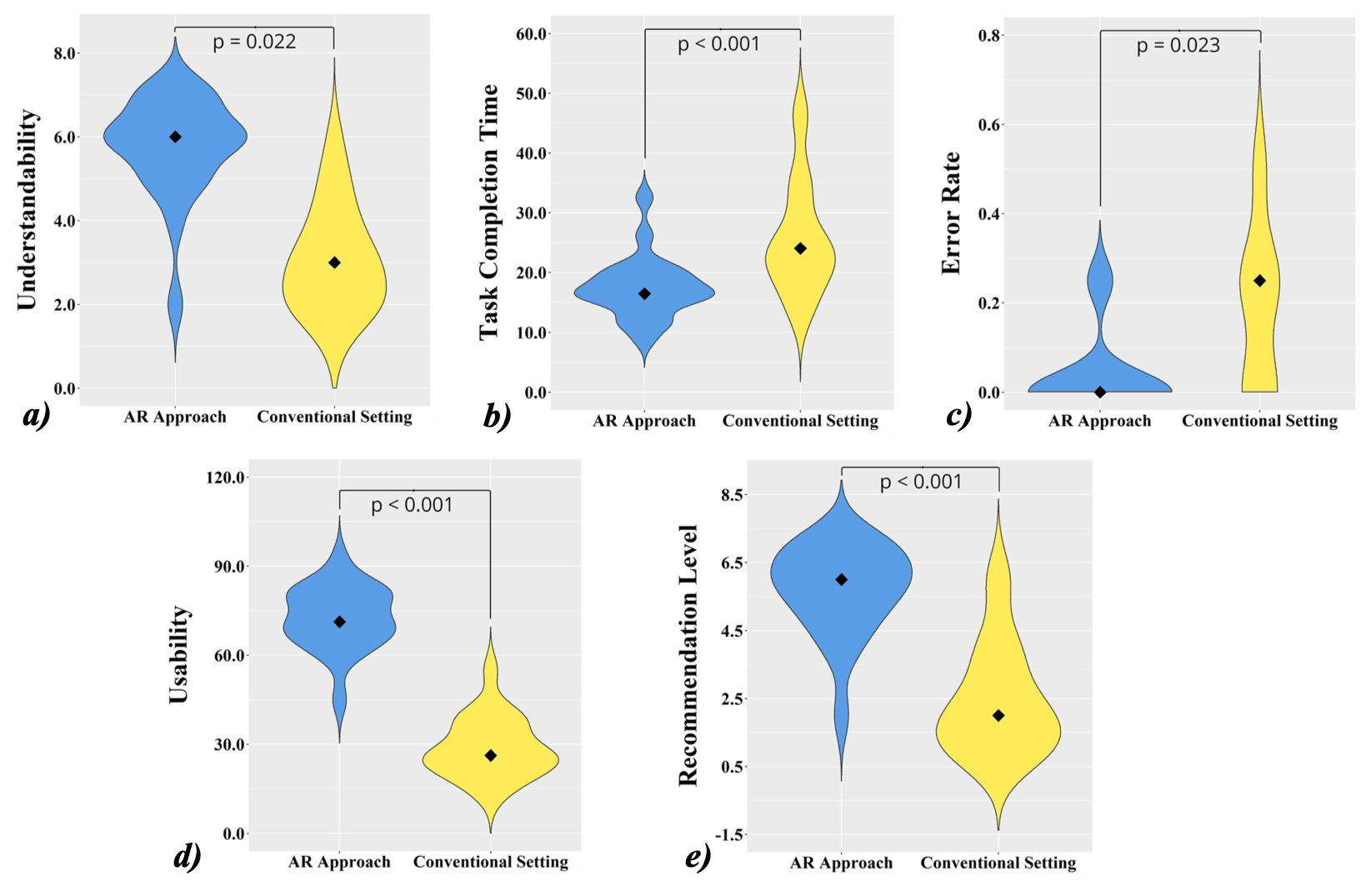}
 \caption{Statistical results between the AR approach and the conventional setting. $a$: The distribution of the understandability levels (1 to 7) rated by participants. $b$: The distribution of task completion time (s). $c$: The distribution of the error rate (0.00 to 1.00). $d$: The distribution of the SUS scores representing usability (0 to 100). $e$: The distribution of the recommendation levels (1 to 7).}
 \label{fig:ues}
\end{figure*}

\subsubsection{Understandability}
This metric was targeted to justify \textbf{H1} which states that the proposed AR approach has a better impact in facilitating IPT visualization analysis. Since the requirement of normality of the collected data did not suffice, a Wilcoxon signed-rank test (non-parametric dependent t-test, confidence interval (CI) 95\%) was then implemented to verify the statistical significance. The result showed that the understandability of the AR approach ($M$=5.75; $SD$=1.36) elicits a statistically significant lead over the conventional setting ($M$=3.00, $SD$=1.41) in IPT contextual understanding ($Z$ = -2.290, $p$ = 0.022). Actually, the median understandability levels were rated as 6 and 3 for AR approach and conventional setting respectively as shown in Figure \ref{fig:ues}.$a$. Also, it shows most of the participants perceived a much higher understandability level with the aid of OST HMD AR in contrast to the conventional settings.

\subsubsection{Task Completion Time (TCT)}
The TCTs in the Understand-n-Select task verified \textbf{H2}. As we elaborated, the reason we only measured the time periods generated from this micro-task was due to them representing the practical IPT visualization task solving. Since the normality of the data was confirmed, a dependent t-test (95\% CI) was implemented which determined that the mean TCT used with the AR approach ($M$=17.56, $SD$=5.31) possessed a statistically significant decline compared to that consumed by the conventional setting ($M$=25.66; $SD$=9.49), ($t$(19) = -3.967, $p$ $<$ 0.001), as shown in Figure \ref{fig:ues}.$b$. The median TCT with the use of AR was 16.47 while that of the conventional setting (24.05) had a noticeable lead (nearly 8s). In other words, participants obtained an obvious performance improvement with our proposed AR approach when conducting IPT related visualization analysis. 

\subsubsection{Error Rate}
The numbers of errors were gathered to investigate \textbf{H2} as well. After the error rate was calculated, we conducted the normality test but it did not show normal distribution on both of the groups. Likewise, we implemented a Wilcoxon signed-rank test with 95\% CI to measure the significance of the two tested scenarios. The result indicated a significant difference between the error rate of the AR approach (M=0.02; SD=0.07) and that of conventional setting ($M$=0.19; $SD$=0.19) with $Z$ = -2.271, $p$ = 0.023. As shown in Figure \ref{fig:ues}.$c$, the distribution plot shows that participants with AR equipment tended to generate much lower error rates when conducting the IPT domain tasks in comparison to being placed in front of the conventional computers. Noteworthily, most of the participants committed 0 errors during the task completion in AR.

\subsubsection{Usability}
\textbf{H3} was designed for examining the usability of our methodology. The fiducial SUS questionnaire was engaged to evaluate the practical usability of the proposed AR approach in helping with IPT visualization analysis. The individual scale values were collected and then calculated to obtain a final SUS score for each participant. A pre-requisite normality testing was implemented and the SUS scores showed normal distribution over the two group conditions. Here, a dependent t-test (95\% CI) determined that the mean usability in the AR approach ($M$=72.71, $SD$=13.12) differed statistically from the usability in the conventional setting ($M$=29.38; $SD$=11.03), ($t$(11) = 6.614, $p$ $<$ 0.001). As illustrated in Figure \ref{fig:ues}.$d$, the median SUS score achieved by the AR approach had an evident lead to conventional setting, implying that satisfactory usability of the proposed AR approach is fully disclosed. In fact, the majority of the participants admitted a substantial exceeding in usability and user experience of the AR method, compared to the conventional settings when tackling IPT contextual visualization analysis.
 
\subsubsection{Recommendation Level}
This metric was harnessed to demonstrate  \textbf{H4}. During the study, each participant was asked to rate how likely they would recommend the two different approaches. As the normality of the collected data was not exhibited, we used a Wilcoxon signed-rank test to identify the statistical significance of the recommendation level. The result indicated that the reported recommendation level from the AR approach ($M$ = 5.70; $SD$ = 1.30) elicits a statistically significant superiority to the conventional setting ($M$ = 2.45, $SD$ = 1.61) in IPT contextual visualization tasks ($Z$ = -3.366, $p$ $<$ 0.001). As Figure \ref{fig:ues}.$e$ shows, the median values of the recommendation level were calculated as 6 and 2 for the AR approach and conventional computer with 2D screen respectively as shown in Figure \ref{fig:ues}. This shows that most of the participants were willing to recommend the OST HMD AR in contrast to the conventional setting regarding IPT related visualization analysis.

\section{Discussion}
\label{dis}

\subsection{Findings}
Through the qualitative and quantitative measurements, we rephrase our main findings as follows. When it refers to IPT related visualization and analysis:

\begin{itemize}
    \item The proposed AR approach with OST HMD can provide better understandability compared to conventional computers with 2D screen.
    \item The proposed AR approach helps users obtain a lower task completion time and fewer domain task errors.
    \item The proposed AR approach can provide better user experience with higher usability regarding interactions.
    \item The proposed AR approach is preferred by users.
\end{itemize}

\subsection{Reflections}
In this study we proposed a framed AR approach to benefit users in the context of IPT visualization analysis. Due the reason that we intended to bring portable AR devices with virtual IPT data visualizations while having the capacity to monitor the ongoing industrial processes in reality, we embraced the AR technique instead of virtual reality (VR).
The proof-of-concept of the AR framework was achieved in HoloLens 2 and the evaluation of a within-subject user study with counterbalancing design was implemented based on the 20 recruited participants to verify the four rasied hypotheses. The qualitative and quantitative resolution demonstrated better understandability for users facing complex IPT data, a lower error rate in related IPT domain task performing, and advanced usability for offering exceptional user experience compared to the conventional setting. 

The total duration of the study fluctuated around 15 minutes, which was advisable and did not cause tiredness of participants in line with our observation. Following \cite{zenner2019estimating,feuchtner2018ownershift}, the "Gorilla arm" effect of fatiguing arms might occur if interacting with tangible/virtual displays for too long, especially in XR environments. We did not consider this as a drawback since our study lasted for an appropriate duration and the effect of the time period on analytical tasks was beyond the scope. However, this factor should be considered as an inspiration for future study design.

Generally, the raw data derived from specialized IPT monitored industrial processes are difficult to interpret \cite{nowak2021augmented}, requiring more advanced tools and techniques for users to conduct in-depth visualization analysis. Our approach offers a complete conceptualization with the realization of an immersive AR method with the aid of OST HMD towards IPT. Our participants were highly favorable towards its interactive features. In particular, the convenient and immersive interactions were highlighted by them to facilitate better comprehension on IPT data, since the conventional settings did not provide the corresponding functions. While no previous work emphasized this, we structured a systematic AR framework with high interaction capabilities for IPT users. In addition, this attribute further facilitated our users' understanding of IPT which resulted in better task performing. Especially, according to Beheiry et al. \cite{el2019virtual}, the immersive experience of VR which makes people feel spatially realistic of 3D data can lead to a "wow" effect by users. We believe that our AR approach brings similar immersion and "wow" effect to users, contributing to high user satisfaction, which is deemed to be an advantage of the proposed methodology since few IPT practitioners have been exposed to AR. Applied to the industrial world, we see the approach enhancing the user experience of a related tooling environment.

Another important property to be realized is the ubiquity and knowledge transferability of the proposed AR approach. It is noteworthy that most of the participants praised the effect of our AR approach of enabling them to understand the context of IPT which resulted in favorable domain task performance. As most of the participants had little experience in IPT, we are therefore encouraged about the potential for our method to bring outsiders to this specialized technique for domain supporting. Additionally, although the data used in this study was generated from a specific IPT -- MWT, the diverse genres of IPT have high transferability since they comply with similar mechanisms and imaging principles \cite{zhang2021supporting}. It is fair to say that the superiority revealed in our AR approach is highly transferable to different genres of IPT.   

\subsection{Limitations}
Even though we obtained early-stage satisfactory results, we have to acknowledge that there still exist some shortcomings. Foremost, the virtual buttons for switching different figures in the AR application were reported as not sensitive enough even if they worked well in this study, particularly due to the limitations of the AR device's distance detecting technology. This could require users to press for longer than they would with physical buttons, causing fatigue and affecting task performance. Additionally, the buttons were designed to be medium in size and positioned on the lower right side of the participants' field of view so that they did not interfere with their manipulation of 3D figures. This could cause the buttons to be relatively small, far, and inconvenient for users to touch. Finally, the lack of haptic feedback may cause confusion when interacting with the virtual buttons, even though visual and aural feedback was provided by the buttons to show whether users had hovered and pressed the button or not. 

To address these issues, future editions of the application could experiment with other button positions, sizes, and feedback mechanisms, as well as incorporate advanced hand gesture recognition technologies. Larger buttons, for example, might be shifted to users' top right side while allowing flexibility for movement by users. Incorporating haptic cues would also improve the user experience when interacting with virtual buttons.

Although hand gestures in AR space are intuitive and highly close to what people commonly do in physical reality, some participants reported a tiny difficulty when the app was initiated. There could be a brief tutorial about basic operation gestures when the participants activate the app, for instance, a short guided video where a pair of virtual hands show how to seize and rotate the virtual objects by pinching fingers and moving arms could be instructive.

Nonetheless, even though the obtained results were satisfying as early-stage feedback, we admit that the experiment itself and the contextual IPT data were monotonous and small scale. To rectify this, the number of participants engaged could be enlarged and more types of experimental design for evaluation could be added. The stimuli in this paper had only three visualizations derived from different industrial processes supervised by the specific MWT, which lacks diversity. More manifold data from other genres of IPT could be examined to make the results more robust.  

\section{Conclusions and Future Work}
\label{cnf}
In this paper, we proposed a novel AR approach with OST HMD for users to immersively interact with the specific IPT data visualizations for contextual understanding and task performing. As the first mechanism of furnishing IPT related users with OST HMD AR, three key findings were explored from our proposed approach. The early-stage advantageous understandability, reduced TCT, lower error rate, greater usability, and higher recommendation level of the methodology were reflected through a within-subject user study. We brought this technique to traditional industrial surroundings, filling the gap between AR developers and IPT domain users. We observed its superiority over the current standard IPT visualization analytical environment, indicating that the immersion in 3D AR outperforms conventional 2D screen computers in enhancing contextual understanding and user experience.

In future, we will firstly concentrate on upgrading the AR app by improving the design of the virtual buttons on the interface, especially focusing on the size design, the touch sensitivity, and the design of the haptic feedback of the buttons as aforementioned. Correspondingly, we will give more early guidance for users to get better adaptation with the app. In addition, another concern will be increasing the experimental diversity by, for instance, involving a larger number of participants including both experts and outsiders and adding more nonidentical experiments with distinct design principles (e.g., between-subject). Promoting inclusivity and diversity of domain users across all races and genders (e.g., engaging non-binary participants) will be at the forefront of our endeavors. Our work benchmarks the intersection between OST HMD AR and IPT to highlight the future direction of bringing more related and advanced techniques into different industrial scenarios. We hope it will lead to better strategic design within this context and bring more interdisciplinary novelty to Industry 4.0 \cite{de2018augmented}.

\subsubsection{Acknowledgements} 
This project has received funding from the European Union’s Horizon 2020 research and innovation programme under the Marie Sklodowska-Curie grant agreement No. 764902. This work was also partly supported by the Norges Forskningsrad (309339, 314578), MediaFutures user partners and Universitetet i Bergen.

%
%
%
\bibliographystyle{splncs04}
\bibliography{mybibliography}

\begin{thebibliography}{10}
\providecommand{\url}[1]{\texttt{#1}}
\providecommand{\urlprefix}{URL }
\providecommand{\doi}[1]{https://doi.org/#1}

\bibitem{alves2022comparing}
Alves, J.B., Marques, B., Ferreira, C., Dias, P., Santos, B.S.: Comparing
  augmented reality visualization methods for assembly procedures. Virtual
  Reality  \textbf{26}(1),  235--248 (2022)

\bibitem{avalle2019augmented}
Avalle, G., De~Pace, F., Fornaro, C., Manuri, F., Sanna, A.: An augmented
  reality system to support fault visualization in industrial robotic tasks.
  IEEE Access  \textbf{7},  132343--132359 (2019)

\bibitem{azuma1997survey}
Azuma, R.T.: A survey of augmented reality. Presence: Teleoperators \& Virtual
  Environments  \textbf{6}(4),  355--385 (1997)

\bibitem{bangor2008empirical}
Bangor, A., Kortum, P.T., Miller, J.T.: An empirical evaluation of the system
  usability scale. Intl. Journal of Human--Computer Interaction
  \textbf{24}(6),  574--594 (2008)

\bibitem{beck2012process}
Beck, M.S., et~al.: Process tomography: principles, techniques and
  applications. Butterworth-Heinemann (2012)

\bibitem{besanccon2017mouse}
Besan{\c{c}}on, L., Issartel, P., Ammi, M., Isenberg, T.: Mouse, tactile, and
  tangible input for 3d manipulation. In: Proceedings of the 2017 CHI
  conference on human factors in computing systems. pp. 4727--4740 (2017)

\bibitem{billinghurst2015survey}
Billinghurst, M., Clark, A., Lee, G., et~al.: A survey of augmented reality.
  Foundations and Trends{\textregistered} in Human--Computer Interaction
  \textbf{8}(2-3),  73--272 (2015)

\bibitem{bottani2019augmented}
Bottani, E., Vignali, G.: Augmented reality technology in the manufacturing
  industry: A review of the last decade. Iise Transactions  \textbf{51}(3),
  284--310 (2019)

\bibitem{bradley1958complete}
Bradley, J.V.: Complete counterbalancing of immediate sequential effects in a
  latin square design. Journal of the American Statistical Association
  \textbf{53}(282),  525--528 (1958)

\bibitem{bruno2006visualization}
Bruno, F., Caruso, F., De~Napoli, L., Muzzupappa, M.: Visualization of
  industrial engineering data visualization of industrial engineering data in
  augmented reality. Journal of visualization  \textbf{9}(3),  319--329 (2006)

\bibitem{buttner2016using}
B{\"u}ttner, S., Funk, M., Sand, O., R{\"o}cker, C.: Using head-mounted
  displays and in-situ projection for assistive systems: A comparison. In:
  Proceedings of the 9th ACM international conference on pervasive technologies
  related to assistive environments. pp.~1--8 (2016)

\bibitem{chen2016using}
Chen, C., Wo{\'z}niak, P.W., Romanowski, A., Obaid, M., Jaworski, T.,
  Kucharski, J., Grudzie{\'n}, K., Zhao, S., Fjeld, M.: Using crowdsourcing for
  scientific analysis of industrial tomographic images. ACM Transactions on
  Intelligent Systems and Technology (TIST)  \textbf{7}(4),  1--25 (2016)

\bibitem{de2018augmented}
De~Pace, F., Manuri, F., Sanna, A.: Augmented reality in industry 4.0. Am J
  ComptSci Inform Technol  \textbf{6}(1), ~17 (2018)

\bibitem{dubois2001consistency}
Dubois, E., Nigay, L., Troccaz, J.: Consistency in augmented reality systems.
  In: IFIP International Conference on Engineering for Human-Computer
  Interaction. pp. 111--122. Springer (2001)

\bibitem{dunn2020stimulating}
Dunn, D., Tursun, O., Yu, H., Didyk, P., Myszkowski, K., Fuchs, H.: Stimulating
  the human visual system beyond real world performance in future augmented
  reality displays. In: 2020 IEEE International Symposium on Mixed and
  Augmented Reality (ISMAR). pp. 90--100. IEEE (2020)

\bibitem{el2019virtual}
El~Beheiry, M., Doutreligne, S., Caporal, C., Ostertag, C., Dahan, M., Masson,
  J.B.: Virtual reality: beyond visualization. Journal of molecular biology
  \textbf{431}(7),  1315--1321 (2019)

\bibitem{ens2021grand}
Ens, B., Bach, B., Cordeil, M., Engelke, U., Serrano, M., Willett, W.,
  Prouzeau, A., Anthes, C., B{\"u}schel, W., Dunne, C., et~al.: Grand
  challenges in immersive analytics. In: Proceedings of the 2021 CHI Conference
  on Human Factors in Computing Systems. pp. 1--17 (2021)

\bibitem{feuchtner2018ownershift}
Feuchtner, T., M{\"u}ller, J.: Ownershift: Facilitating overhead interaction in
  virtual reality with an ownership-preserving hand space shift. In:
  Proceedings of the 31st Annual ACM Symposium on User Interface Software and
  Technology. pp. 31--43 (2018)

\bibitem{fite2011there}
Fite-Georgel, P.: Is there a reality in industrial augmented reality? In: 2011
  10th ieee international symposium on mixed and augmented reality. pp.
  201--210. IEEE (2011)

\bibitem{hampel2022review}
Hampel, U., Babout, L., Banasiak, R., Schleicher, E., Soleimani, M., Wondrak,
  T., Vauhkonen, M., L{\"a}hivaara, T., Tan, C., Hoyle, B., et~al.: A review on
  fast tomographic imaging techniques and their potential application in
  industrial process control. Sensors  \textbf{22}(6), ~2309 (2022)

\bibitem{heemsbergen2021conceptualising}
Heemsbergen, L., Bowtell, G., Vincent, J.: Conceptualising augmented reality:
  From virtual divides to mediated dynamics. Convergence  \textbf{27}(3),
  830--846 (2021)

\bibitem{henderson2011augmented}
Henderson, S.J., Feiner, S.K.: Augmented reality in the psychomotor phase of a
  procedural task. In: 2011 10th IEEE international symposium on mixed and
  augmented reality. pp. 191--200. IEEE (2011)

\bibitem{ismail2005tomography}
Ismail, I., Gamio, J., Bukhari, S.A., Yang, W.: Tomography for multi-phase flow
  measurement in the oil industry. Flow measurement and instrumentation
  \textbf{16}(2-3),  145--155 (2005)

\bibitem{jasche2021comparison}
Jasche, F., Hoffmann, S., Ludwig, T., Wulf, V.: Comparison of different types
  of augmented reality visualizations for instructions. In: Proceedings of the
  2021 CHI Conference on Human Factors in Computing Systems. pp. 1--13 (2021)

\bibitem{kahl2021investigation}
Kahl, D., Ruble, M., Kr{\"u}ger, A.: Investigation of size variations in
  optical see-through tangible augmented reality. In: 2021 IEEE International
  Symposium on Mixed and Augmented Reality (ISMAR). pp. 147--155. IEEE (2021)

\bibitem{kalkofen2008comprehensible}
Kalkofen, D., Mendez, E., Schmalstieg, D.: Comprehensible visualization for
  augmented reality. IEEE transactions on visualization and computer graphics
  \textbf{15}(2),  193--204 (2008)

\bibitem{khan2021measuring}
Khan, F.A., Rao, V.V.R.M.K., Wu, D., Arefin, M.S., Phillips, N., Swan, J.E.,
  et~al.: Measuring the perceived three-dimensional location of virtual objects
  in optical see-through augmented reality. In: 2021 IEEE International
  Symposium on Mixed and Augmented Reality (ISMAR). pp. 109--117. IEEE (2021)

\bibitem{kress2014segmentation}
Kress, B., Saeedi, E., Brac-de-la Perriere, V.: The segmentation of the hmd
  market: optics for smart glasses, smart eyewear, ar and vr headsets.
  Photonics Applications for Aviation, Aerospace, Commercial, and Harsh
  Environments V  \textbf{9202},  107--120 (2014)

\bibitem{leebmann2004augmented}
Leebmann, J.: An augmented reality system for earthquake disaster response.
  International Archives of the Photogrammetry, Remote Sensing and Spatial
  Information Sciences  \textbf{34}(Part XXX) (2004)

\bibitem{liu112018application}
Liu11, X., Sohn, Y.H., Park, D.W.: Application development with augmented
  reality technique using unity 3d and vuforia. International Journal of
  Applied Engineering Research  \textbf{13}(21),  15068--15071 (2018)

\bibitem{lorenz2018industrial}
Lorenz, M., Knopp, S., Klimant, P.: Industrial augmented reality: Requirements
  for an augmented reality maintenance worker support system. In: 2018 IEEE
  International Symposium on Mixed and Augmented Reality Adjunct
  (ISMAR-Adjunct). pp. 151--153. IEEE (2018)

\bibitem{mann2001augmented}
Mann, R., Stanley, S., Vlaev, D., Wabo, E., Primrose, K.: Augmented-reality
  visualization of fluid mixing in stirred chemical reactors using electrical
  resistance tomography. Journal of Electronic Imaging  \textbf{10}(3),
  620--630 (2001)

\bibitem{masood2020adopting}
Masood, T., Egger, J.: Adopting augmented reality in the age of industrial
  digitalisation. Computers in Industry  \textbf{115},  103112 (2020)

\bibitem{mourtzis2020augmented}
Mourtzis, D., Siatras, V., Zogopoulos, V.: Augmented reality visualization of
  production scheduling and monitoring. Procedia CIRP  \textbf{88},  151--156
  (2020)

\bibitem{nolet2012seismic}
Nolet, G.: Seismic tomography: with applications in global seismology and
  exploration geophysics, vol.~5. Springer Science \& Business Media (2012)

\bibitem{nowak2019towards}
Nowak, A., Wo{\'z}niak, M., Rowi{\'n}ska, Z., Grudzie{\'n}, K., Romanowski, A.:
  Towards in-situ process tomography data processing using augmented reality
  technology. In: Adjunct Proceedings of the 2019 ACM International Joint
  Conference on Pervasive and Ubiquitous Computing and Proceedings of the 2019
  ACM International Symposium on Wearable Computers. pp. 168--171 (2019)

\bibitem{nowak2021augmented}
Nowak, A., Zhang, Y., Romanowski, A., Fjeld, M.: Augmented reality with
  industrial process tomography: To support complex data analysis in 3d space.
  In: Adjunct Proceedings of the 2021 ACM International Joint Conference on
  Pervasive and Ubiquitous Computing and Proceedings of the 2021 ACM
  International Symposium on Wearable Computers. pp. 56--58 (2021)

\bibitem{omrani2022multistatic}
Omrani, A., Yadav, R., Link, G., Jelonnek, J.: A multistatic uniform
  diffraction tomography algorithm for microwave imaging in multilayered media
  for microwave drying. IEEE Transactions on Antennas and Propagation  (2022)

\bibitem{ong2013virtual}
Ong, S.K., Nee, A.Y.C.: Virtual and augmented reality applications in
  manufacturing. Springer Science \& Business Media (2013)

\bibitem{peillard2020can}
Peillard, E., Itoh, Y., Moreau, G., Normand, J.M., L{\'e}cuyer, A., Argelaguet,
  F.: Can retinal projection displays improve spatial perception in augmented
  reality? In: 2020 IEEE International Symposium on Mixed and Augmented Reality
  (ISMAR). pp. 80--89. IEEE (2020)

\bibitem{pierdicca2015making}
Pierdicca, R., Frontoni, E., Zingaretti, P., Malinverni, E.S., Colosi, F.,
  Orazi, R.: Making visible the invisible. augmented reality visualization for
  3d reconstructions of archaeological sites. In: International Conference on
  Augmented and Virtual Reality. pp. 25--37. Springer (2015)

\bibitem{plaskowski1995imaging}
Plaskowski, A., Beck, M., Thorn, R., Dyakowski, T.: Imaging industrial flows:
  applications of electrical process tomography. CRC Press (1995)

\bibitem{primrose2015application}
Primrose, K.: Application of process tomography in nuclear waste processing.
  In: Industrial Tomography, pp. 713--725. Elsevier (2015)

\bibitem{rao2022monitoring}
Rao, G., Aghajanian, S., Zhang, Y., Strumillo, L.J., Koiranen, T., Fjeld, M.:
  Monitoring and visualization of crystallization processes using electrical
  resistance tomography: Caco3 and sucrose crystallization case studies  (2022)

\bibitem{regenbrecht2005augmented}
Regenbrecht, H., Baratoff, G., Wilke, W.: Augmented reality projects in the
  automotive and aerospace industries. IEEE computer graphics and applications
  \textbf{25}(6),  48--56 (2005)

\bibitem{satkowski2021investigating}
Satkowski, M., Dachselt, R.: Investigating the impact of real-world
  environments on the perception of 2d visualizations in augmented reality. In:
  Proceedings of the 2021 CHI Conference on Human Factors in Computing Systems.
  pp. 1--15 (2021)

\bibitem{sobiech2022exploratory}
Sobiech, F., Walczak, N., Buczek, A., Jeanty, M., Kupi{\'n}ski, K., Chaniecki,
  Z., Romanowski, A., Grudzie{\'n}, K.: Exploratory analysis of users’
  interactions with ar data visualisation in industrial and neutral
  environments  (2022)

\bibitem{de2020survey}
de~Souza~Cardoso, L.F., Mariano, F.C.M.Q., Zorzal, E.R.: A survey of industrial
  augmented reality. Computers \& Industrial Engineering  \textbf{139},  106159
  (2020)

\bibitem{speicher2019mixed}
Speicher, M., Hall, B.D., Nebeling, M.: What is mixed reality? In: Proceedings
  of the 2019 CHI conference on human factors in computing systems. pp. 1--15
  (2019)

\bibitem{stanley2005interrogation}
Stanley, S., Mann, R., Primrose, K.: Interrogation of a precipitation reaction
  by electrical resistance tomography (ert). AIChE journal  \textbf{51}(2),
  607--614 (2005)

\bibitem{tainaka2020guideline}
Tainaka, K., Fujimoto, Y., Kanbara, M., Kato, H., Moteki, A., Kuraki, K.,
  Osamura, K., Yoshitake, T., Fukuoka, T.: Guideline and tool for designing an
  assembly task support system using augmented reality. In: 2020 IEEE
  International Symposium on Mixed and Augmented Reality (ISMAR). pp. 486--497.
  IEEE (2020)

\bibitem{tapp2003chemical}
Tapp, H., Peyton, A., Kemsley, E., Wilson, R.: Chemical engineering
  applications of electrical process tomography. Sensors and Actuators B:
  Chemical  \textbf{92}(1-2),  17--24 (2003)

\bibitem{tonnis2005experimental}
Tonnis, M., Sandor, C., Klinker, G., Lange, C., Bubb, H.: Experimental
  evaluation of an augmented reality visualization for directing a car driver's
  attention. In: Fourth IEEE and ACM International Symposium on Mixed and
  Augmented Reality (ISMAR'05). pp. 56--59. IEEE (2005)

\bibitem{yao2017application}
Yao, J., Takei, M.: Application of process tomography to multiphase flow
  measurement in industrial and biomedical fields: A review. IEEE Sensors
  Journal  \textbf{17}(24),  8196--8205 (2017)

\bibitem{zenner2019estimating}
Zenner, A., Kr{\"u}ger, A.: Estimating detection thresholds for desktop-scale
  hand redirection in virtual reality. In: 2019 IEEE Conference on Virtual
  Reality and 3D User Interfaces (VR). pp. 47--55. IEEE (2019)

\bibitem{zhang2020automated}
Zhang, Y., Ma, Y., Omrani, A., et~al.: Automated microwave tomography (mwt)
  image segmentation: State-of-the-art implementation and evaluation. Journal
  of WSCG  \textbf{2020},  126--136 (2020)

\bibitem{zhang2020condition}
Zhang, Y., Fjeld, M.: Condition monitoring for confined industrial process
  based on infrared images by using deep neural network and variants. In:
  Proceedings of the 2020 2nd International Conference on Image, Video and
  Signal Processing. pp. 99--106 (2020)

\bibitem{zhang2021affective}
Zhang, Y., Fjeld, M., Fratarcangeli, M., Said, A., Zhao, S.: Affective colormap
  design for accurate visual comprehension in industrial tomography. Sensors
  \textbf{21}(14), ~4766 (2021)

\bibitem{zhang2023need}
Zhang, Y., Nowak, A., Rao, G., Romanowski, A., Fjeld, M.: Is industrial
  tomography ready for augmented reality? a need-finding study of how augmented
  reality can be adopted by industrial tomography experts. In: International
  Conference on Human-Computer Interaction. Springer (2023, in press)

\bibitem{zhang2022initial}
Zhang, Y., Nowak, A., Romanowski, A., Fjeld, M.: An initial exploration of
  visual cues in head-mounted display augmented reality for book searching. In:
  Proceedings of the 21st International Conference on Mobile and Ubiquitous
  Multimedia. pp. 273--275 (2022)

\bibitem{zhang2022site}
Zhang, Y., Nowak, A., Romanowski, A., Fjeld, M.: On-site or remote working?: An
  initial solution on how covid-19 pandemic may impact augmented reality users.
  In: Proceedings of the 2022 International Conference on Advanced Visual
  Interfaces. pp.~1--3 (2022)

\bibitem{zhang2021supporting}
Zhang, Y., Omrani, A., Yadav, R., Fjeld, M.: Supporting visualization analysis
  in industrial process tomography by using augmented reality—a case study of
  an industrial microwave drying system. Sensors  \textbf{21}(19), ~6515 (2021)

\bibitem{zhang2021novel}
Zhang, Y., Yadav, R., Omrani, A., Fjeld, M.: A novel augmented reality system
  to support volumetric visualization in industrial process tomography. In:
  Proceedings of the 2021 Conference on Interfaces and Human Computer
  Interaction. pp.~3--9 (2021)

\end{thebibliography}
%




\end{document}